\documentclass[twocolumn]{IEEEtran} 

\usepackage{amsmath,amssymb,epsfig}

\title{On the stable recovery of the sparsest overcomplete representations in presence of noise} 
\author{
  \thanks{Copyright \copyright\ 2010 IEEE. Personal use of this material is permitted. However, permission to use this material for any other purposes must be obtained from the IEEE by sending a request to pubs-permissions@ieee.org.}
  Massoud Babaie-Zadeh,~\IEEEmembership{Senior Member},
  Christian Jutten,~\IEEEmembership{Fellow}
  \thanks{M. Babaie-Zadeh is with the Electrical engineering department, Sharif University of
    Technology, Tehran, Iran (email: {\tt mbzadeh@yahoo.com}).} 
  \thanks{C. Jutten is with the GIPSA-Lab, University of Grenoble, and the Institut Universitaire de France, France (email: {\tt Christian.Jutten@gipsa-lab.grenoble-inp.fr}).}  
  \thanks{This work has been partially funded by the Iran National Science Foundation (INSF), by the Iran Telecom Research Center (ITRC), and also by the Center for International Research and Collaboration (ISMO) and the French embassy in Tehran in the framework of a GundiShapour collaboration program.}}

\newcommand{\nb}{{\bf n}}

\newcommand{\Ab}{{\bf A}}

\newcommand{\Bb}{{\bf B}}

\newcommand{\ab}{{\bf a}}

\newcommand{\xb}{{\bf x}}

\newcommand{\sbb}{{\bf s}}
\newcommand{\hsbb}{{\bf \hat s}}
\newcommand{\hxb}{{\bf \hat x}}

\newcommand{\vb}{{\bf v}}

\newcommand{\Phib}{{\mbox{\boldmath {$\Phi$ }}}}
\newcommand{\Psib}{{\mbox{\boldmath {$\Psi$ }}}}
\newcommand{\ie}{{\em i.e.}}
\newcommand{\eg}{{\em e.g.}}

\newcommand{\spark}{\mbox{\sl{spark}}}

\newcommand{\sm}[1]{\sigma_{\textrm{\rm min}}^{(#1)}}

\newcommand{\nz}[1]{\|#1\|_0} 
\newcommand{\no}[1]{\|#1\|_1} 
\newcommand{\nt}[1]{\|#1\|_2} 
\newcommand{\eps}{\varepsilon}

\newtheorem{theorem}{Theorem}

\begin{document}
\maketitle

\begin{abstract}
Let $\xb$ be a signal to be sparsely decomposed over a redundant dictionary $\Ab$, \ie\, a sparse coefficient vector $\sbb$ has to be found such that $\xb=\Ab\sbb$. It is known that this problem is inherently unstable against noise, and to overcome this instability, the authors of 
\cite{DonoET06}
have proposed to use an ``approximate'' decomposition, that is, a decomposition satisfying $\|{\bf x} - {\bf A s}\| \le \delta$ rather than satisfying the exact equality ${\bf x} = {\bf A s}$. 
Then, they have shown that if there is a decomposition with $\|{\bf s}\|_0 < (1+M^{-1})/2$, where $M$ denotes the {\em coherence} of the dictionary, this decomposition would be stable against noise. On the other hand, it is known that a sparse decomposition with $\|{\bf s}\|_0 < \frac{1}{2} \mbox{\sl spark}({\bf A})$ is unique. In other words, although a decomposition with $\|{\bf s}\|_0 < \frac{1}{2} \mbox{\sl spark}({\bf A})$ is unique, its stability against noise has been proved only for highly more restrictive decompositions satisfying $\|{\bf s}\|_0 < (1+M^{-1})/2$, because usually $(1+M^{-1})/2 \ll \frac{1}{2} \mbox{\sl spark}({\bf A})$.

This limitation maybe had not been very important before, because $\|{\bf s}\|_0 < (1+M^{-1})/2$ is also the bound which guaranties that the sparse decomposition can be found via minimizing the $\ell^1$ norm, a classic approach for sparse decomposition. However, with the availability of new algorithms for sparse decomposition, namely SL0 and Robust-SL0, it would be important to know whether or not unique sparse decompositions with $(1+M^{-1})/2 \le \|{\bf s}\|_0 < \frac{1}{2} \mbox{\sl spark}({\bf A})$ are stable. In this paper, we show that such decompositions are indeed stable. In other words, we extend the stability bound from $\|{\bf s}\|_0 < (1+M^{-1})/2$ to the whole uniqueness range $\|{\bf s}\|_0 < \frac{1}{2} \mbox{\sl spark}({\bf A})$. In summary, we show that {\em all unique sparse decompositions are stably recoverable}. Moreover, we see that sparser decompositions are `more stable'.
\end{abstract}

\begin{IEEEkeywords}
Sparse Signal Decomposition, Sparse recovery, Compressed Sensing, Sparse Component Analysis (SCA), Overcomplete dictionaries.
\end{IEEEkeywords}

\section{Introduction}
\IEEEPARstart{L}{et} $\Ab$ be an $n \times m$ matrix with $m>n$, and consider the Underdetermined System of Linear Equations (USLE) $\Ab \sbb = \xb$. Such a linear system has typically infinitely many solutions, but let consider its sparsest solution, that is, a solution $\sbb_0$ which has as much as possible zero components.

This problem has recently attracted a lot of attention from many different viewpoints. It is used, for example, in Compressed Sensing (CS)~\cite{CandRT06,Dono06,Bara07}, underdetermined Sparse Component Analysis (SCA) and source separation~\cite{GribL06, BofiZ01, GeorTC04, LiCA03}, atomic decomposition on overcomplete dictionaries~\cite{ChenDS99,DonoET06},  decoding real field codes~\cite{CandT05}, image deconvolution~\cite{FiguN03,FiguN05}, image denoising~\cite{Elad06}, electromagnetic imaging and Direction of Arrival (DOA) finding~\cite{GoroR97}, etc. 

In atomic decomposition viewpoint~\cite{MallZ93}, the columns of $\Ab$ are called `atoms' and the matrix $\Ab$ is called the `dictionary' over which the `signal' $\xb$ is to be decomposed. When the dictionary is overcomplete ($m>n$), the representation is not unique, but by the sparsest solution, we are looking for the representation which uses as small as possible number of atoms to represent the signal.

Sparse solutions of underdetermined linear systems would not be useful, unless positive answers can be provided for the following three questions:
\begin{enumerate}
	\item {\bf Uniqueness:} Is such a solution unique?
	
	\item {\bf Practical algorithm:} Is it practically possible to find the sparsest solution of an USLE?
	
	\item {\bf Stability against noise:} Doesn't a small amount of noise result in a completely different sparse solution?
\end{enumerate}

In this paper we study the third question, and we generalize previously available results. To better explain the problem and our contribution, we firstly do a brief review in Section~\ref{sec: Problem statement} on the available results about the above questions, and then explain in subsection~\ref{sec: our contrib} what our contribution is. We state then the main theorem in Section~\ref{sec: Main Thm}. Finally, a generalized result will be stated in Section~\ref{sec: Gen Thm}.

\section{Problem statement}\label{sec: Problem statement}
\subsection{Uniqueness?}
The uniqueness problem has been addressed in~\cite{GoroR97,GribN03,DonoE03}, and it has been shown that if an underdetermined linear system has a sparse enough solution, it would be its unique sparsest solution. More precisely:

\medskip
\begin{theorem}[Uniqueness~\cite{GribN03,DonoE03}]
\label{th: uniqueness}\em
Let $\spark(\Ab)$ denote the minimum number of columns of $\Ab$ that are linearly dependent, and $\|\cdot\|_0$ denotes the $\ell^0$ norm of a vector (i.e. the number of its non-zero components). Then if the USLE $\Ab \sbb = \xb$ has a solution $\sbb_0$ for which $\|\sbb_0\|_0 < \frac{1}{2} \spark(\Ab)$, it is its unique sparsest solution.
\end{theorem}
\medskip

A special case of this uniqueness theorem has been stated in~\cite{GoroR97}: if $\Ab$ has the Unique Representation Property (URP), that is, if all $n \times n$ submatrices of $\Ab$ are non-singular, then $\spark(\Ab)=n+1$ and hence $\|\sbb_0\|_0 \le \frac{n}{2}$ implies that $\sbb_0$ is the unique sparsest solution.

\subsection{Practical Algorithm?}
Finding the sparsest solution of an USLE can be expressed as: 
\begin{equation}
	(P_0): \quad \mbox{Minimize } \nz{\sbb} \quad \mbox{subject to}\quad \Ab \sbb = \xb,
\end{equation}
where $\nz{\cdot}$ stands for the $\ell^0$ norm of a vector. Solving the above problem requires a combinatorial search and is generally NP-hard. Then, many algorithms have been proposed to indirectly solve the problem. One of the first and most successful ideas is the idea of Basis Pursuit (BP)~\cite{ChenDS99}, which is to replace the above problem by
\begin{equation}
	(P_1): \quad \mbox{Minimize } \no{\sbb} \quad \mbox{subject to}\quad \Ab \sbb = \xb,
\end{equation}
where $\no{\sbb}\triangleq\sum_i|s_i|$ is the $\ell^1$ norm of \sbb.
Note that the problem $P_1$ is convex and can be easily solved by using Linear Programming (LP) techniques. Moreover, it has been shown that if the sparsest solution $\sbb_0$ is highly sparse, then the solution of $P_1$ is also the sparsest solution, \ie\ it is also the solution of $P_0$. 

To express this property more precisely, let the columns of $\Ab$ be normalized to have unit $\ell^2$ (Euclidean) norm. Let also define the {\em `coherence'}, $M$, of the dictionary $\Ab$ as the maximum correlation between its atoms, that is:
\begin{equation}
	M \triangleq \max_{i\neq j} |\ab_i^T \ab_j|,
\end{equation}
where $\ab_i$, $i=1,...,m$ denote the columns of $\Ab$. Then:

\medskip
\begin{theorem}[Equivalence of $P_0$ and $P_1$~\cite{GribN03,DonoE03}] 
\label{th: equiv P0 P1}\em
If the USLE $\Ab \sbb = \xb$ has a solution $\sbb_0$ for which $\|\sbb_0\|_0 < \frac{1+M^{-1}}{2}$, then it is the unique solution of both problems $P_0$ and $P_1$.
\end{theorem}
\medskip

In other words, if the sparsest solution satisfies $\|\sbb_0\|_0 < \frac{1+M^{-1}}{2}$, it can be found by solving the convex program $P_1$.

\medskip

{\bf Remark 1.\ } Note that the bound on sparsity that guaranties the equivalence of $P_0$ and $P_1$ is highly more restrictive than the bound which guaranties the uniqueness of the sparsest solution. For example, suppose that the dictionary $\Ab$ is constructed by concatenating two orthonormal bases, $\Ab=[\Phib, \Psib]$, and hence $m=2n$. It can be easily shown~\cite{GribN03} that in this case the maximum possible value for $M$ is $1/\sqrt{n}$ (this maximum value for $M$ is obtained for example for concatenation of a Dirac and a Fourier dictionary). Consider for example such a dictionary $\Ab$ with $m=1000$ and $n=500$, which satisfies the URP and has the maximum possible coherence $M=\frac{1}{\sqrt{n}}\approx {1}/{(22.36)}$. Then, by Theorem~\ref{th: uniqueness} a solution $\sbb_0$ with $\nz{\sbb_0} \le 250$ is necessarily the unique sparsest solution. However, from Theorem~\ref{th: equiv P0 P1}, it is guaranteed that the sparsest solution can be found by $P_1$ only where $\nz{\sbb_0} < (1+22.36)/2$, that is $\nz{\sbb_0} \le 11$. In other words, if there is a solution $\sbb_0$ such that among its 1000 entries there are at most 250 non-zero entries, it would be the unique sparsest solution, but we cannot necessarily find it by solving $P_1$, unless among these 1000 entries, there are at most 11 non-zero entries. Consequently, equivalence of $P_1$ and $P_0$ holds only for the case there exists a {\em `very very' sparse} solution.

\medskip

{\bf Remark 2.\ } Note also that if the unique sparsest solution satisfies $\frac{1+M^{-1}}{2} \le \|\sbb_0\|_0 < \frac{1}{2} \spark(\Ab)$, the above theorem does not state that it `cannot' be found by solving $P_1$; it simply does not `guarantee' that $P_1$ can recover it. In fact, from the uniqueness Theorem~\ref{th: uniqueness}, we know that if we find a solution $\hsbb_0$ by using any method (\eg\ $P_1$, or even simply by a magic guess), and we see that it happens that $\|\hsbb_0\|_0 < \frac{1}{2} \spark(\Ab)$, we will know that we have found the unique sparsest solution.

\medskip

In addition to the methods based on $\ell^1$ norm minimization, there are other ideas for finding the sparsest solution, for example Matching Pursuit (MP)~\cite{MallZ93} and Smoothed $\ell^0$ (SL0)~\cite{MohiBJ09}. The latter method (SL0), which has been designed in our group, tries to directly solve the $P_0$ problem by replacing the $\ell^0$ norm by a smooth approximation of it (and hence the name `smoothed' $\ell^0$). One of the motivations behind SL0 is the fact stated above: Since the equivalence of $P_0$ and $P_1$ holds only where there exist very very sparse solutions, it would probably be better trying to solve $P_0$ directly. Another motivation is the speed: it has been shown~\cite{MohiBJ09} that SL0 is highly faster than solving $P_1$.

\subsection{Stability against noise?} \label{sec: stability}
Suppose that $\xb_0$ is a linear combination of a few atoms of the dictionary, that is, $\xb_0=\Ab \sbb_0$, where $\sbb_0$ is sparse. Now consider a noisy measurement of $\xb_0$, that is, $\xb=\xb_0+ \nb$, where $\nb$ denotes the noise, and $\nt{\nb} \le \eps$. 
The question of `stability'~\cite{DonoET06} is then:
Even for a very small $\eps$, is it guaranteed that the sparse decomposition of $\xb$ over the dictionary (problem $P_0$) is not too different from the sparse decomposition of $\xb_0$? 
The answer is unfortunately no, that is, the problem $P_0$ can be too sensitive to noise~\cite{Wohl03}.

To overcome this problem, it has been proposed in~\cite{DonoET06} that instead of solving $P_0$ or $P_1$ one considers solving their noise aware variants:
\begin{gather}
(P_{0,\delta}): \quad \mbox{Minimize } \nz{\sbb} \quad \mbox{s.t.}\quad \nt{\xb - \Ab \sbb} \le \delta \\
(P_{1,\delta}): \quad \mbox{Minimize } \no{\sbb} \quad \mbox{s.t.}\quad \nt{\xb - \Ab \sbb} \le \delta
\end{gather}
In other words, it has been proposed to do an ``approximate'' decomposition, that is, a decomposition with $\|{\bf x} - {\bf A s}\| \le \delta$ instead of the exact decomposition ${\bf x} = {\bf A s}$.
These noise aware variants have to be solved for a sufficiently large $\delta$, that is, for $\delta \ge \eps$ to guarantee that the true solution $\sbb_0$ satisfies the constraints of the above optimization problems. Then, in~\cite{DonoET06}, the authors prove that both problems $P_{0,\delta}$ and $P_{1,\delta}$ are stable against noise, that is, the estimation error is at worst proportional to the noise level. More precisely, the stability of $P_{0,\delta}$ is given by the following theorem:

\medskip
\begin{theorem}[Stability of $P_{0,\delta}$; theorem 2.1 of \cite{DonoET06}] \label{th: Donoho}
\em
Let $M$ denote the coherence of the dictionary $\Ab$. Suppose that for the sparse representation of the noiseless signal $\xb_0=\Ab \sbb_0$ we have:
\begin{equation}
	k \triangleq \nz{\sbb_0} < \frac{1+M^{-1}}{2}
	\label{eq: Condition Classic}
\end{equation}
If $\hsbb_{0,\delta}$ denotes the result of applying $P_{0,\delta}$ on the noisy data $\xb$ with $\delta \ge \eps$, then:
\begin{equation}
	\nt{\hsbb_{0,\delta}-\sbb_0} \le \frac{\eps+\delta}{\sqrt{1-M(2k-1)}}\, .
	\label{eq: error bound classic}
\end{equation}
\end{theorem}
\bigskip

Note that (\ref{eq: Condition Classic}) implies also that the term under the square root in (\ref{eq: error bound classic}) is positive.

The authors of \cite{DonoET06} also prove the stability of $P_{1,\delta}$ for the case $\nz{\sbb_0} < {(1+M^{-1})}/{4}$.

A noise aware variant of SL0 (called Robust-SL0), has already been developed~\cite{EfteBJA09}, which tries to solve directly $P_{0,\delta}$ without passing through $P_{1,\delta}$.

\subsection{Our Contribution}\label{sec: our contrib}
As it was said in Section~\ref{sec: stability}, the stability of the problem $P_{0,\delta}$ has only been shown for the case $\nz{\sbb_0} < {(1+M^{-1})}/{2}$. This sparsity limit for stability is the same as the sparsity limit for the equivalence of $P_0$ and $P_1$ as stated in Theorem~\ref{th: equiv P0 P1}. However, as was stated in Remark~1 after Theorem~\ref{th: equiv P0 P1}, this sparsity limit is highly more restrictive than the sparsity limit for the uniqueness of the sparse solution.
In other words, current results state that although a sparse representation with $\frac{1+M^{-1}}{2} \le \nz{\sbb_0} < \frac{1}{2} \spark(\Ab)$ is unique, it is not guaranteed that $P_{0,\delta}$ can stably recover this representation in presence of noise. 

Maybe the lack of this guarantee had not been important before, because, the classic idea for solving $P_0$ was solving $P_1$, and the sparsity limit for the equivalence of these two solutions is the same as the sparsity limit for the stability of $P_{0,\delta}$. However, with new algorithms like SL0 or Robust-SL0, one can now try to solve $P_{0,\delta}$ directly and without relying on $P_{1,\delta}$. Hence it is now important to know whether or not sparse representations with $\frac{1+M^{-1}}{2} \le \nz{\sbb_0} < \frac{1}{2} \spark(\Ab)$ are stable.

In the next section, we will show that $P_{0,\delta}$ is stable for the {\em whole sparsity range} that guarantees the uniqueness, that is, $P_{0,\delta}$ is stable whenever $\nz{\sbb_0} < \frac{1}{2} \spark(\Ab)$. Moreover, we will show that for smaller $\nz{\sbb_0}$ the problem is `more stable', that is, {\em the more sparsity, the more stability}. Finally, we will show in Section~\ref{sec: Gen Thm} that {\em this stability not only holds for $P_{0,\delta}$, but also holds for any estimation $\hsbb_0$ such that $\nz{\hsbb_0} < \frac{1}{2} \spark(\Ab)$ and $\nt{\xb - \Ab \hsbb_0} \le \delta$}.

\section{The main theorem}\label{sec: Main Thm}
To state the main theorems, we need first to define two notations:

\begin{itemize}
	\item Let $q=q(\Ab)=\spark(\Ab)-1$. Then, by definition, every $q$ columns of $\Ab$ are linearly independent, and there is at least a set of $q+1$ columns which are linearly dependent (in the literature, the quantity $q$ is usually called `Kruskal rank' or `k-rank' of the matrix $\Ab$). It is also obvious that $q\le n$, in which, $q=n$ corresponds to the case $\Ab$ has the URP.
	
	\item Let $\sm{j}$, $1\le j \le q(\Ab)$, denote the smallest singular value among all of the submatrices of $\Ab$ formed by taking $j$ columns of $\Ab$. Note that since every $q$ columns of $\Ab$ are linearly independent, we have $\sm{j} > 0$, $\forall 1\le j \le q(\Ab)$.
\end{itemize}

Moreover, it is known~\cite[p. 419]{HornJ85},~\cite[Lemma 3]{BabaMJ09} that if we add a new column to a full-rank tall matrix, its smallest singular value decreases or remains the same (refer to \cite{BabaMJ09} for a simple direct proof). Therefore, $\sm{j}$ is a decreasing sequence in $j$, that is:
\begin{equation}\label{eq: sigma decrease}
	\sm{j} \ge \sm{j+1} > 0, \quad \forall 1 \le j \le q-1
\end{equation}

We are now ready to state the following theorem.

\medskip
\begin{theorem}[Stability of $P_{0,\delta}$]
\label{th: Main Theorem}\em
Suppose that the noiseless signal $\xb_0$ has a sparse representation $\xb_0=\Ab \sbb_0$ satisfying $\nz{\sbb_0} <
\frac{1}{2}\spark(\Ab)$. Let also $\xb=\xb_0+\nb$ be a noisy measurement of $\xb_0$ and $\nt{\nb} \le \eps$.
If $\hsbb_{0,\delta}$ denotes the result of applying $P_{0,\delta}$ on the noisy signal $\xb$ with $\delta \ge \eps$, then:
\begin{equation}
	\nt{\hsbb_{0,\delta}-\sbb_0} \le \frac{\delta+\eps}{\sm{\ell}},
	\label{eq: our first upper bound}
\end{equation}
where $\ell = 2 \nz{\sbb_0}$.
\end{theorem}
\medskip

{\bf Remark~1.\ } Theorem~\ref{th: Main Theorem} shows that $P_{0,\delta}$ is stable not only for $\nz{\sbb_0} < \frac{1+M^{-1}}{2}$, but also for the whole uniqueness range $\nz{\sbb_0} < \frac{1}{2} \spark(\Ab)$. The stability is in the sense that the estimation error increases at worst proportionally to the noise level. Moreover, from (\ref{eq: sigma decrease}), the upper bound on estimation error decreases or remains the same as the sparsity increases (this is because sparser $\sbb_0$ means smaller $\nz{\sbb_0}$, which implies smaller $\ell$ and hence larger or the same $\sm{\ell}$). In other words, {\em sparser solutions are `more stable'}.

\medskip

{\bf Remark~2.\ } The main reason for stating Theorem~\ref{th: Main Theorem} is to provide a stability result for the case $1+M^{-1} \le \ell=2 \nz{\sbb_0} < \spark(\Ab)$, because in this case, Theorem~\ref{th: Donoho} provides 
no stability result. Moreover, note that for the case $\ell < 1+M^{-1}$, in which both bounds (\ref{eq: error bound classic}) and (\ref{eq: our first upper bound}) are applicable, (\ref{eq: our first upper bound}) provides also a tighter bound than (\ref{eq: error bound classic}). This is implied from Lemma~2.2 of~\cite{DonoET06} which states that in this case $\sm{\ell}>\sqrt{1-M(\ell - 1)}$. 

\medskip

\begin{IEEEproof}[Proof of Theorem~\ref{th: Main Theorem}]
Let define $\hxb_{0,\delta} \triangleq \Ab \hsbb_{0,\delta}$. We write:
\begin{align}
 	  \nt{\xb_0 - \hxb_{0,\delta}} &= \nt{\xb-\nb - \hxb_{0,\delta}} \nonumber \\
 	  &= \nt{(\xb - \Ab \hsbb_{0,\delta}) -\nb} \nonumber \\
 	  &\le \underbrace{\nt{\xb - \Ab \hsbb_{0,\delta}}}_{\le \delta} + \underbrace{\nt{\nb}}_{\le \eps} \nonumber \\
 	  &\le \delta+\eps \label{eq: upper bound on x diff}
\end{align}

\noindent On the other hand:
\begin{equation}
\label{eq: temp1}
	\xb_0 - \hxb_{0,\delta} = \Ab (\sbb_0 - \hsbb_{0,\delta}) = \Bb \vb
\end{equation}
where $\vb$ is a vector composed of non-zero entries of $\sbb_0 - \hsbb_{0,\delta}$, and 
$\Bb$ is a submatrix of $\Ab$ composed of the columns of $\Ab$ corresponding to the non-zero entries of $\sbb_0 - \hsbb_{0,\delta}$.
Since $\delta \ge \eps$, $\sbb_0$ satisfies the constraint of the optimization problem $P_{0, \delta}$, and hence $\nz{\hsbb_{0,\delta}} \le \nz{\sbb_0}$. Therefore $\sbb_0 - \hsbb_{0,\delta}$ has at most $\ell\triangleq2 \nz{\sbb_0}<\spark(\Ab)$ non-zero entries (note that $\ell < \spark(\Ab)$ means $\ell \le q(\Ab)$). In other words, $\Bb$ has at most $\ell\le q$ columns, and hence (by having also in mind (\ref{eq: sigma decrease})):
\begin{equation}
	\nt{\Bb \vb} \ge \sm{\ell} \nt{\vb}
	\label{eq: Bv low bound}
\end{equation}
Noting that $\nt{\vb}=\nt{\sbb_0 - \hsbb_{0,\delta}}$, and combining the above inequality with (\ref{eq: temp1}), we obtain:
\begin{equation} \label{eq: lower bound on x diff}
	\nt{\xb_0 - \hxb_{0,\delta}} \ge \sm{\ell} \nt{\sbb_0 - \hsbb_{0,\delta}}
\end{equation}
Combining (\ref{eq: upper bound on x diff}) and (\ref{eq: lower bound on x diff}) gives:
\begin{equation}
	\sm{\ell} \nt{\sbb_0 - \hsbb_{0,\delta}} \le \delta + \eps
\end{equation}
which completes the proof.
\end{IEEEproof}
\medskip

{\bf Remark~3.\ } From (\ref{eq: sigma decrease}) and $\ell=2 \nz{\sbb_0} \le q(\Ab)$, we may replace $\sm{j}$ by its worst case to obtain the following looser bound, which does not need knowing the value of $\nz{\sbb_0}$:
\begin{equation}
	\nt{\hsbb_{0,\delta}-\sbb_0} \le \frac{\delta+\eps}{\sm{q}}.
	\label{eq: looser bound}
\end{equation}

\section{A generalized stability theorem}\label{sec: Gen Thm}
If we carefully re-examine the proof of Theorem~\ref{th: Main Theorem}, we notice that the fact that $\nz{\hsbb_{0,\delta}}\le\nz{\sbb_0}$ is not essential for obtaining the looser bound~(\ref{eq: looser bound}). Hence, the bound~(\ref{eq: looser bound}) holds not only for the sparse recovery methods based on solving $P_{0, \delta}$, but also for {\em any other estimation} $\hsbb_{0,\delta}$ (obtained from any sparse recovery algorithm or even simply from a magic guess), 
provided that it satisfies $\nz{\hsbb_{0,\delta}} < \frac{1}{2} \spark(\Ab)$ and $\nt{\xb - \Ab\hsbb_{0,\delta}} \le \delta$. 
In other words, {\em not only $P_{0,\delta}$ is stable, but also any other method for `approximate' sparse representation is stable\/} provided that it provides a sparse enough estimation.
More precisely:

\medskip
\begin{theorem}[Stability of approximate sparse representation]
\label{th: Generalized Theorem}\em
Suppose that the noiseless signal $\xb_0$ has a sparse representation $\xb_0=\Ab \sbb_0$ satisfying $\nz{\sbb_0} <
\frac{1}{2}\spark(\Ab)$. Let also $\xb=\xb_0+\nb$ be a noisy measurement of $\xb_0$ and $\nt{\nb} \le \eps$.
If we have at hand an estimation $\hsbb_{0,\delta}$ of the sparse representation coefficients which satisfies $\nz{\hsbb_{0,\delta}} < \frac{1}{2} \spark(\Ab)$ and $\nt{\xb - \Ab\hsbb_{0,\delta}} \le \delta$, then:
\begin{equation}
	\nt{\hsbb_{0,\delta}-\sbb_0} \le \frac{\delta+\eps}{\sm{q}},
	\label{eq: gen bound}
\end{equation}
\end{theorem}
\medskip

\begin{IEEEproof} It is easily obtained by following the same steps as the proof of Theorem~\ref{th: Main Theorem}: equations (\ref{eq: upper bound on x diff}) and (\ref{eq: temp1}) still hold. We then note that:
\begin{equation}
	\nz{\hsbb_{0,\delta}-\sbb_0} \le \nz{\hsbb_{0,\delta}} + \nz{\sbb_0} < \spark(\Ab)
\end{equation}
and hence $\nz{\hsbb_{0,\delta}-\sbb_0}\le q(\Ab)$. Consequently, instead of (\ref{eq: Bv low bound}) we write:
\begin{equation}
   \nt{\Bb \vb} \ge \sm{q} \nt{\vb}	
\end{equation}
which in combination by (\ref{eq: upper bound on x diff}) and (\ref{eq: temp1}) proves (\ref{eq: gen bound}).
\end{IEEEproof}

\medskip

{\bf Remark.\ } Note that the condition $\delta \ge \eps$ does not explicitly appeared in Theorem~\ref{th: Generalized Theorem}, and is no more essential (while it was essential in Theorem~\ref{th: Main Theorem}, because it was necessary to insure that $P_{0,\delta}$ gives an estimation satisfying $\nz{\hsbb_{0,\delta}}\le\nz{\sbb_0}$, which was essential in the proof). However, implicitly, the $\delta$ in Theorem~\ref{th: Generalized Theorem} cannot be too small, because for a very small $\delta$, it is possible that there exists no $\hsbb_{0,\delta}$ satisfying $\nt{\xb-\Ab\hsbb_{0,\delta}}\le \delta$.

\section{Conclusion}
Since minimizing $\ell^1$ norm has been one of the first and most successful ideas for finding the sparsest solution of an USLE, some theoretical aspects of the sparsest solution are currently too much influenced by the $\ell^1$ minimization idea. Currently, with the availability of the algorithms that try to find the sparse solution by means of other approaches, \eg\ SL0 and Robust-SL0, some of the properties of the sparsest solution need to be revisited. In this paper, we studied the stability of the sparsest solution, and we showed that it is stable not only where $\nz{\sbb_0}< (1+M^{-1})/2$, but also for the whole uniqueness range $\nz{\sbb_0}< \frac{1}{2} \spark(\Ab)$. These results prove the practical interest of designing $\ell^0$-norm minimization algorithms, since they can provide a good estimation from noisy data, with the weakest condition of sparsity.

\bibliography{sepsrc}

\begin{thebibliography}{10}

\bibitem{DonoET06}
D.~L. Donoho, M.~Elad, and V.~Temlyakov,
\newblock ``Stable recovery of sparse overcomplete representations in the
  presence of noise,''
\newblock {\em IEEE Trans. Info. Theory}, vol. 52, no. 1, pp. 6--18, Jan 2006.

\bibitem{CandRT06}
E.J. Cand\`{e}s, J.~Romberg, and T.~Tao,
\newblock ``Robust uncertainty principles: exact signal reconstruction from
  highly incomplete frequency information,''
\newblock {\em IEEE Transactions on Information Theory}, vol. 52, no. 2, pp.
  489--509, February 2006.

\bibitem{Dono06}
D.~L. Donoho,
\newblock ``Compressed sensing,''
\newblock {\em IEEE Transactions on Information Theory}, vol. 52, no. 4, pp.
  1289--1306, April 2006.

\bibitem{Bara07}
R.~G. Baraniuk,
\newblock ``Compressive sensing,''
\newblock {\em IEEE Signal Processing Magazine}, vol. 24, no. 4, pp. 118--124,
  July 2007.

\bibitem{GribL06}
R.~Gribonval and S.~Lesage,
\newblock ``A survey of sparse component analysis for blind source separation:
  principles, perspectives, and new challenges,''
\newblock in {\em Proceedings of ESANN'06}, April 2006, pp. 323--330.

\bibitem{BofiZ01}
P.~Bofill and M.~Zibulevsky,
\newblock ``Underdetermined blind source separation using sparse
  representations,''
\newblock {\em Signal Processing}, vol. 81, pp. 2353--2362, 2001.

\bibitem{GeorTC04}
P.~G. Georgiev, F.~J. Theis, and A.~Cichocki,
\newblock ``Blind source separation and sparse component analysis for
  over-complete mixtures,''
\newblock in {\em Proceedinds of ICASSP'04}, Montreal (Canada), May 2004, pp.
  493--496.

\bibitem{LiCA03}
Y.~Li, A.~Cichocki, and S.~Amari,
\newblock ``Sparse component analysis for blind source separation with less
  sensors than sources,''
\newblock in {\em ICA2003}, 2003, pp. 89--94.

\bibitem{ChenDS99}
S.~S. Chen, D.~L. Donoho, and M.~A. Saunders,
\newblock ``Atomic decomposition by basis pursuit,''
\newblock {\em SIAM Journal on Scientific Computing}, vol. 20, no. 1, pp.
  33--61, 1999.

\bibitem{CandT05}
E.~J. Cand\`{e}s and T.~Tao,
\newblock ``Decoding by linear programming,''
\newblock {\em IEEE Transactions on Information Theory}, vol. 51, no. 12, pp.
  4203--4215, 2005.

\bibitem{FiguN03}
M.~A.~T. Figueiredo and R.~D. Nowak,
\newblock ``An {EM} algorithm for wavelet-based image restoration,''
\newblock {\em IEEE Transactions on Image Processing}, vol. 12, no. 8, pp.
  906--916, 2003.

\bibitem{FiguN05}
M.~A.~T. Figueiredo and R.~D. Nowak,
\newblock ``A bound optimization approach to wavelet-based image
  deconvolution,''
\newblock in {\em IEEE Internation Conference on Image Processing (ICIP)},
  August 2005, pp. II--782--5.

\bibitem{Elad06}
M.~Elad,
\newblock ``Why simple shrinkage is still relevant for redundant
  representations?,''
\newblock {\em IEEE Transactions on Image Processing}, vol. 52, no. 12, pp.
  5559--5569, 2006.

\bibitem{GoroR97}
I.~F. Gorodnitsky and B.~D. Rao,
\newblock ``Sparse signal reconstruction from limited data using {FOCUSS}, a
  re-weighted minimum norm algorithm,''
\newblock {\em IEEE Transactions on Signal Processing}, vol. 45, no. 3, pp.
  600--616, March 1997.

\bibitem{MallZ93}
S.~Mallat and Z.~Zhang,
\newblock ``Matching pursuits with time-frequency dictionaries,''
\newblock {\em IEEE Trans. on Signal Proc.}, vol. 41, no. 12, pp. 3397--3415,
  1993.

\bibitem{GribN03}
R.~Gribonval and M.~Nielsen,
\newblock ``Sparse decompositions in unions of bases,''
\newblock {\em IEEE Trans. Inform. Theory}, vol. 49, no. 12, pp. 3320--3325,
  Dec. 2003.

\bibitem{DonoE03}
D.~L. Donoho and M.~Elad,
\newblock ``Optimally sparse representation in general (nonorthogonal)
  dictionaries via $\ell^1$ minimization,''
\newblock {\em Proc. Nat. Aca. Sci.}, vol. 100, no. 5, pp. 2197--2202, March
  2003.

\bibitem{MohiBJ09}
H.~Mohimani, M.~Babaie-Zadeh, and Ch. Jutten,
\newblock ``A fast approach for overcomplete sparse decomposition based on
  smoothed $\ell^0$ norm,''
\newblock {\em IEEE Transactions on Signal Processing}, vol. 57, no. 1, pp.
  289--301, January 2009.

\bibitem{Wohl03}
B.~Wohlberg,
\newblock ``Noise sensitivity of sparse signal representations: Reconstruction
  error bounds for the inverse problem,''
\newblock {\em IEEE Transaction on Signal Processing}, vol. 51, no. 12, pp.
  3053--3060, December 2003.

\bibitem{EfteBJA09}
A.~Eftekhari, M.~Babaie-Zadeh, Ch. Jutten, and H.~Abrishami-Moghaddam,
\newblock ``Robust-{SL0} for stable sparse representation in noisy settings,''
\newblock in {\em Proceedings of {\em ICASSP2009}}, Taipei, Taiwan, 19--24
  April 2009, pp. 3433--3436.

\bibitem{HornJ85}
R.~A. Horn and C.~R. Johnson,
\newblock {\em Matrix analysis},
\newblock Cambridge University Press, Cambridge, 1985.

\bibitem{BabaMJ09}
M.~Babaie-Zadeh, H.~Mohimani, and Ch. Jutten,
\newblock ``An upper bound on the estimation error of the sparsest solution of
  underdetermined linear systems,''
\newblock in {\em Proceedings of {\em SPARS2009}}, Saint-Malo, France, 6--9
  April 2009.

\end{thebibliography}
\bibliographystyle{IEEEbib}

\begin{biography}[{\includegraphics[width=1in,height=1.25in,clip,keepaspectratio]{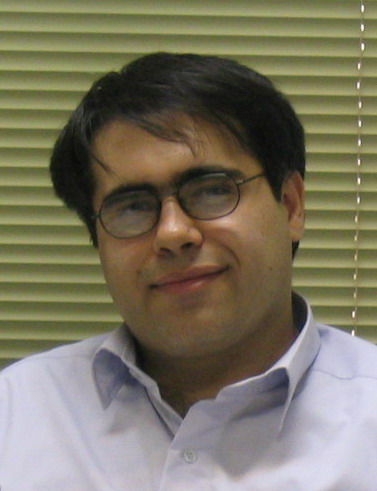}}]%
{{\bf Massoud Babaie-Zadeh} \rm (M'04-SM'09) received the B.S. degree in electrical engineering from Isfahan University of Technology, Isfahan, Iran in 1994, and the M.S degree in electrical engineering from Sharif University of Technology, Tehran, Iran, in 1996, and the Ph.D degree in Signal Processing from Institute National Polytechnique of Grenoble (INPG), Grenoble, France, in 2002.\\
\hspace*{1.5em}Since 2003, he has been a faculty member of the Electrical Engineering Department of Sharif University of Technology, Tehran, IRAN, firstly as an assistant professor and since 2008 as an associate professor. His main research areas are Blind Source Separation (BSS) and Independent Component Analysis (ICA), Sparse Signal Processing, and Statistical Signal Processing.\\
\hspace*{1.5em}Dr. Babaie-Zadeh received the best Ph.D. thesis award of INPG for his Ph.D. dissertation.
}
\end{biography}

\begin{biography}[{\includegraphics[width=1in,height=1.25in,clip,keepaspectratio]{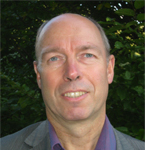}}]%
{{\bf Christian Jutten} \rm (AM'92-M'03-SM'06-F'08) received the PhD degree in 1981 and the Docteur ès Sciences degree in 1987 from the Institut National Polytechnique of Grenoble (France). After being associate professor in the Electrical Engineering Department (1982-1989) and visiting professor in Swiss Federal Polytechnic Institute in Lausanne (1989), he became full professor in University Joseph Fourier of Grenoble, more precisely in the sciences and technologies department. For 30 years, his research interests are learning in neural networks, blind source separation and independent component analysis, including theoretical aspects (separability, source separation in nonlinear mixtures) and applications (biomedical, seismic, speech).  He is author or co-author of more than 55 papers in international journals, 4 books, 18 invited papers and 160 communications in international conferences. He was co-organizer of the 1st International Conference on Blind Signal Separation and Independent Component Analysis (Aussois, France, January 1999). He has been a scientific advisor for signal and images processing at the French Ministry of Research (1996-1998) and for the French National Research Center (2003-2006). He has been associate editor of IEEE Trans. on Circuits and Systems (1994-95). He is a member of the technical committee "Blind signal Processing" of the IEEE CAS society and of the technical committee "Machine Learning for signal Processing" of the IEEE SP society. He received the EURASIP best paper award in 1992 and Medal Blondel in 1997 from SEE (French Electrical Engineering society) for his contributions in source separation and independent component analysis, and has been elevated as a Fellow IEEE and a senior Member of Institut Universitaire de France in 2008.
}
\end{biography}

\end{document}